\documentclass[
aps,
prd,
showpacs,
tightenlines,
superscriptaddress,
nofootinbib,
floatfix]
{revtex4}
\usepackage{graphicx,
longtable
}

\usepackage{epsfig}
\begin{document}
\title{		Combined analysis of KamLAND and Borexino neutrino signals\\ 
           from Th and U decays in the Earth's interior}
\author{			G.L.\ Fogli}
\affiliation{Dipartimento Interateneo di Fisica ``Michelangelo Merlin,'' \\ Via Amendola 173, 70126 Bari, Italy}
\affiliation{INFN, Sezione di Bari, \\ Via Orabona 4, 70126, Bari, Italy}
\author{	          E.\ Lisi}
\affiliation{INFN, Sezione di Bari, \\ Via Orabona 4, 70126, Bari, Italy}
\author{			A.\ Palazzo}
\affiliation{AHEP Group, Institut de F\'isica Corpuscular -- C.S.I.C./Universitat de Val\'encia,\\ 
             Edificio Institutos de Paterna, P.O. Box 22085, EÐ46071 Valencia, Spain}
\author{			A.M.\ Rotunno}
\affiliation{Dipartimento Interateneo di Fisica ``Michelangelo Merlin,'' \\ Via Amendola 173, 70126 Bari, Italy}
\affiliation{INFN, Sezione di Bari, \\ Via Orabona 4, 70126, Bari, Italy}


\begin{abstract}
The KamLAND and Borexino experiments have detected  
electron antineutrinos produced in the decay chains of natural thorium and uranium (Th and U geoneutrinos). 
We analyze the energy spectra of current geoneutrino data in combination with solar and long-baseline reactor 
neutrino data, with marginalized three-neutrino oscillation parameters. We consider
the case with unconstrained Th and U  event rates in KamLAND and Borexino, as well as cases with fewer 
degrees of freedom, as obtained by successively assuming for both experiments
a common Th/U ratio, a common scaling of $\mathrm{Th}+\mathrm{U}$ event rates, and a chondritic Th/U value.
In combination, KamLAND and Borexino can reject the null hypothesis (no geoneutrino signal) at $5\sigma$.  
Interesting bounds or indications emerge on the $\mathrm{Th}+\mathrm{U}$ geoneutrino rates and on 
the Th/U ratio, in broad agreement with typical Earth model expectations. 
Conversely, the results disfavor
the hypothesis of a georeactor in the Earth's core, if its power exceeds
a few TW. The interplay of KamLAND and Borexino 
geoneutrino data is highlighted. 
\end{abstract}
 \maketitle


\section{Context}

The study of low-energy electron antineutrinos ($\overline\nu_e$) emitted in the decay chains of
uranium (U), thorium (Th), and potassium (K) in the Earth's interior---the so-called 
geoneutrinos---is raising 
increasing interest in both particle physics and Earth sciences, as recently reviewed in \cite{Fi07,Mc08}.

From the viewpoint of particle physics, there has been dramatic progress in understanding 
the flavor evolution and oscillations of neutrinos \cite{Fo06}, and in refining their  low-energy,
low-background detection techniques. In particular, the detection of the global 
$\mathrm{Th}+\mathrm{U}$ geoneutrino flux in the KamLAND \cite{KL05,KL08} and Borexino \cite{BX10} experiments
through the inverse beta decay reaction 
\begin{equation}
\label{IBD} \overline\nu_e+ p\to n + e^+ \ (E_\nu>1.806\ \mathrm{MeV})\ ,
\end{equation}
represents a milestone in this field.%
\footnote{Geo-$\nu$ from K decay are below threshold for the reaction (\ref{IBD}).}
In perspective, measurements at different locations might constrain the   
relative Th and U abundances in different reservoirs (e.g., crust versus mantle), 
especially if some directional sensitivity could 
be achieved \cite{Fi07}. 
 
From the viewpoint of Earth sciences, the heat-producing elements U, Th and K, despite their relatively low 
natural abundances, bear on outstanding and debated issues. Their global amounts should reflect different 
condensation histories in  the primitive solar nebula, which presumably led to 
partial escape of the moderately ``volatile'' K and to
complete capture of the ``refractory''  U and Th. Therefore, the mass abundance ratio of Th and U
in the Earth (Th/U hereafter) is expected to be the same as in the most pristine meteorite samples,
the so-called carbonaceous chondrites \cite{Mc08},
\begin{equation}
\label{CHO}
\frac{\mathrm{Th}}{\mathrm{U}}\simeq 3.9 \ (\mathrm{chondritic\ estimate})\ .
\end{equation}
The radial distribution of U, Th and K should instead reflect subsequent Earth differentiation processes, 
as these elements are both 
``lithophile'' (preferring mantle and crust silicates to core metals)
and ``incompatible'' (preferring crust melts to mantle residues). Within the mantle,
slow convection processes may have further redistributed the radiogenic elements in
several possible ways \cite{Mc08}.

In this field, connecting geophysical quantities to particle physics observables 
requires some modeling of the U, Th, and K distributions, on both planetary and local scales. 
For instance, a correlation is expected between the radiogenic heat production rate $H$ 
and the geoneutrino event rate $R$ from $\mathrm{Th}+\mathrm{U}$ sources, within 
large uncertainties induced, e.g., by the unknown distribution
of radiogenic elements in the mantle. Various $(H,\,R)$ correlation plots have been discussed in \cite{Fi07} under 
rather general and plausible assumptions [including the estimate in~(\ref{CHO})] 
and conservative uncertainties. In particular, the results of \cite{Fi07} for KamLAND
(see Figs.~23 and 30 therein) can be approximated as 
\begin{equation}
\label{HR}
\frac{H(\mathrm{Th}+\mathrm{U})}{\mathrm{TW}} \simeq (1.11\pm0.14)\times \frac{R(\mathrm{Th}+\mathrm{U})}{\mathrm{TNU}}-25.0  
\ \ \mathrm{(KamLAND},\ 6~\mathrm{TW}\lesssim H\lesssim 40~\mathrm{TW)} \ , 
\end{equation}
where 1~TNU (terrestrial neutrino unit) corresponds to $10^{32}$~events per target proton per year, and
the quoted error provides a sort of ``maximum allowed range'' for the $(H,\,R)$ correlation band. The above
estimate holds for $H\gtrsim 6$~TW, which provides a total KamLAND signal in excess of the ``guaranteed'' 
minimum contribution from the Earth's crust: $R(\mathrm{Th}+\mathrm{U})\gtrsim 24$~TNU \cite{Fi07}. On the other
hand, Eq.~(\ref{HR}) is not applicable beyond the ``fully radiogenic'' limit $H(\mathrm{Th}+\mathrm{U})\simeq  40$~TW \cite{Fi07},
which, adding an estimated potassium contribution $H(\mathrm{K})\simeq 5$~TW \cite{Mc09}, 
would saturate the global Earth's heat flow, $H_\oplus\simeq 45$~TW \cite{Po93,Ja07}.

Positive correlations are also expected among the expected event rates in different experiments,
since they probe the same geoneutrino sources, although 
weighted differently by the inverse square law for the fluxes. 
In particular, it is rather plausible to assume that KamLAND (KL) and Borexino (BX) probe the same
average Th/U ratio, so that
\begin{equation}
\label{THU}
\frac{R(\mathrm{U})_\mathrm{BX}}{R(\mathrm{Th})_\mathrm{BX}}\simeq 
\frac{R(\mathrm{U})_\mathrm{KL}}{R(\mathrm{Th})_\mathrm{KL}}\ .
\end{equation}
Even if the primordial proportions of Th and U in the Earth were different from the estimate in~(\ref{CHO}), the
known geochemical similarity of Th and U in different reservoirs would support the above assumption. 

Correlations among absolute rates (rather than ratio of rates) may be more model-dependent.
For instance, 
a comparison of the KamLAND (KL) and Borexino (BX) uranium event rates estimated in \cite{Fi07} for a wide range of admissible
Earth models (see Table~12 and Fig.~24 therein) suggests an approximate scaling law,
\begin{equation}
\label{SCALE}
R(\mathrm{U})_\mathrm{BX}\simeq 1.15\, R(\mathrm{U})_\mathrm{KL}\ , 
\end{equation}
within a relatively small spread ($< 5\%$) in the coefficient. A scaling coefficient $>1$ is to be expected,
since Borexino probes a thicker crust than KamLAND. Its spread, however, may actually be larger than $5\%$, since
the local (and sizable) crust contributions at the two sites are not expected to   
have significant covariances, thus reducing the overall correlation \cite{Fo07}. 
Nevertheless, within the large uncertainties affecting current geoneutrino event rates, 
an approximate scaling assumption as in Eq.~(\ref{SCALE}) can still provide a useful guidance in the data
analysis, and will be used later.

Summarizing, the most general analysis of 
the available KamLAND and Borexino data involves four geoneutrino degrees of freedom ($N_D=4$), namely, the thorium
and uranium event rates in the two experiments. This parameter space can be reduced 
by assuming either of the two Eqs.~(\ref{THU},\ref{SCALE}). If the estimate in~(\ref{CHO}) is also imposed, a single
degree of freedom remains $(N_D=1)$. The main purpose of this work is to analyze geoneutrino data, and to
discuss their implications, in 
cases with $N_D=4$, 3, 2, and 1. A special case is also discussed, where an additional
degree of freedom is provided by the unknown power of a hypothetical georeactor in the Earth's core \cite{He96}. 

Our paper is structured as follows. In Section~II we discuss our approach to the geoneutrino analysis, with particular attention to 
the theoretical and experimental energy spectra and their uncertainties. 
Since low-energy neutrino oscillations---constrained by
solar and long-baseline reactor data---affect the extraction of geoneutrino signals, 
the marginalization of the three-neutrino oscillation parameters in a global fit is also discussed.
In Section~III we show the results of our analysis in four relevant cases ($N_D=4$, 3, 2, and 1),
and discuss their implications. In general, we find results in agreement
with typical Earth model expectations, although often within large uncertainties. An interesting
interplay between KamLAND and Borexino data emerges in all cases with $N_D\leq 3$. Finally,
we show that the georeactor hypothesis is disfavored by the data. 
Our conclusions are summarized in Section~IV.

\section{Method}

In this Section we describe some aspects of our analysis, concerning 
the geoneutrino and  reactor energy spectra, 
the marginalization of oscillation parameters, the input data from KamLAND and Borexino, and the 
relevant geoneutrino degrees of freedom under increasingly restrictive assumptions about the relative event rates.

\subsection{Geoneutrino energy spectra}

\begin{figure}[t]
\centering
\epsfig{figure=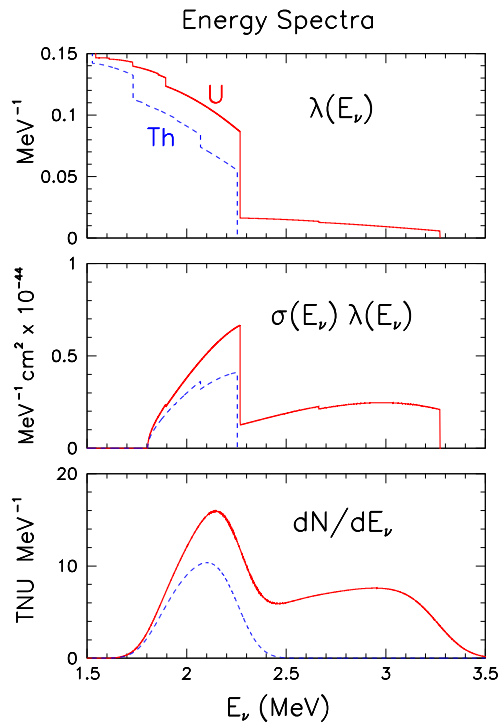,width=0.36\columnwidth}
\caption{Geoneutrino spectra computed in this work, as a function
of the neutrino energy $E_\nu$. Upper panel: spectra from  U and Th decay chains (solid and dashed curves, 
respectively). Middle panel: spectra
multiplied by the cross section for inverse beta decay. Lower panel: observable event spectra,
including typical energy resolution effects. The results refer to 
a $\overline\nu_e$ flux of $10^6$/cm$^2$/s (all panels), and to $10^{32}$ target protons (lower panel).
\label{fig_01}}
\end{figure}

The reaction (\ref{IBD}) allows to detect (a fraction of) the $4~\overline\nu_e$ and the $6~\overline\nu_e$ 
produced in the decay chain
of the $^{232}$Th and the $^{238}$U nuclei, respectively,
as described at length in \cite{Fi07}, to which we refer the reader for details. 
We perform our own calculation of the corresponding energy spectra,
based on experimental nuclear data from \cite{Isot} and theoretical inputs from \cite{BeJa}.

Our results are reported in Fig.~1.  The upper panel shows the computed energy spectra $\lambda(E)$ from the Th and U decay chains,
assuming in both cases a reference $\overline\nu_e$ flux of $10^6$/cm$^2$/s. The spectra
are normalized to unit area for $E_\nu\geq 0$ (lower range not shown). From our spectra
we estimate that the flux fraction above the 1.806~MeV threshold is 0.151/4 (Th) and 0.384/6 (U). 
The middle panel of Fig.~1 shows the spectra times the inverse beta decay cross section \cite{StVi}.
We estimate the average cross sections as $\sigma_\mathrm{Th}=12.9\times 10^{-46}$~cm$^2$ and 
$\sigma_\mathrm{U}=40.5\times 10^{-46}$~cm$^2$, corresponding to the event rates $R(\mathrm{Th})=4.07$~TNU
and $R(\mathrm{U})=12.8$~TNU for a reference $\overline\nu_e$ flux of $10^6$/cm$^2$/s.
Since 1~kg of natural Th
emits $16.2\times 10^6$ $\overline\nu_e$/s from $^{232}$Th decays, and 1~kg of natural U emits 
$74.1\times 10^6$ $\overline\nu_e$/s from $^{238}$U decays \cite{Fi07}, the event rates and 
the natural mass abundances of Th and U for a given source are related by 
\begin{equation}
\label{CONV}
\frac{R(\mathrm{Th})}{R(\mathrm{U})}=6.96\times 10^{-2}\,\frac{\mathrm{Th}}{\mathrm{U}}\ .
\end{equation}

The lower panel of Fig.~1 shows 
the typical effect of finite energy resolution in liquid-scintillator  detectors such as KamLAND and Borexino.
The event rate spectra appear to be
significantly smeared out, in the range $E_\nu~\simeq 1.7$--3.5~MeV. Separation of Th and U contribution requires,
in principle, an accurate determination of
relative event rates above and below $E_\nu~\simeq 2.5$~MeV. Experimentally, one does not measure $E_\nu$
but the observable ``prompt'' energy $E_p$
associated to the final-state positron and its annihilation \cite{KL05,KL08,BX10},  
\begin{equation}
E_p = T_{e^+} + 2m_e =  E_\nu - (m_n-m_p)+ m_e \simeq E_\nu - 0.782~\mathrm{MeV}\ .
\end{equation}
Accounting for smearing effects on $E_p$, approximate geoneutrino energy windows for Th and U events are then 
 $E_p(\mathrm{Th})\in[0.9,\, 1.7]$~MeV and
$E_p(\mathrm{U})\in[0.9,\, 2.6]$~MeV, respectively.

\subsection{Reactor energy spectra}

Concerning KamLAND, we have described our approach to the
reactor spectra and data analysis in previous works \cite{Fo06,Prev1,Prev2,Prev3}, to which we refer the 
reader for details. Concerning Borexino, we integrate available
information on position, type and average power of European and world reactors from several 
public sources, see e.g.\ \cite{Nucl},
and adopt the typical power fractions of fuel components as suggested in \cite{BX10}. Since the oscillated reactor
spectra at Borexino are largely averaged over time and over many (and very long) baselines---besides
being smeared by energy resolution effects---more accurate
information is not really needed for our purposes.

\subsection{Marginalization of oscillation parameters}

In KamLAND and Borexino, geoneutrino events can be distinguished by reactor and background events
only statistically, on the basis of their different energy spectra.
The reactor spectra are significantly affected by three-neutrino 
oscillations  governed by the squared mass gap
$\delta m^2=m^2_2-m^2_1$ and  by the mixing angles 
$\theta_{12}$ and $\theta_{13}$ \cite{Fo06}. The pattern of $\delta m^2$-driven oscillations clearly emerges
in KamLAND reactor spectra after pathlengths $L\sim O(10^2)$~km \cite{KL08}, while the pattern is 
largely averaged out in the case of Borexino, where $L\sim O(10^3)$~km \cite{BX10}. 
Complete averaging of oscillations can be assumed, to a good approximation, for geoneutrinos \cite{Fi07}.

The statistical separation of geoneutrino, reactor and background spectra in KamLAND and Borexino
depends thus on the  oscillation parameters 
$(\delta m^2,\,\theta_{12},\theta_{13})$ which, in turn, are also constrained by solar
neutrino data \cite{Fo06}. We perform a combined analysis of KamLAND and Borexino  data,
together with all solar neutrino data, updating our previous work \cite{Prev3,Hint}.
In particular, we include the latest Gallium experiment event
rates \cite{SAGE,GALL} and the low-energy threshold data from the Sudbury Neutrino Observatory (SNO) \cite{LETA}. We also consider
both low- and high-metallicity options for the Standard Solar Model \cite{Sere}. 
The statistical analysis involves a 7-dimensional manifold, spanned by
\begin{equation}
\label{PAR} 
\big\{\delta m^2,\,\theta_{12},\,\theta_{13}
;\,R(\mathrm{Th})_\mathrm{KL}
,\,R(\mathrm{U})_\mathrm{KL}
,\,R(\mathrm{Th})_\mathrm{BX}
,\,R(\mathrm{U})_\mathrm{BX}
\big\} ,
\end{equation}
plus a number of nuisance parameters which account for systematic uncertainties via the
pull method \cite{Pull} (see also the next subsection). 

In the global data fit, the
marginalization of the four geoneutrino rates $R$ provides bounds on the low-energy oscillation parameters 
$(\delta m^2,\theta_{12},\theta_{13})$, which will be discussed elsewhere. In this work we are
interested in the complementary case, where the marginalization of  
$(\delta m^2,\theta_{12},\theta_{13})$ provides constraints on the Th and U event rates in KamLAND and Borexino.

\subsection{KamLAND and Borexino data}

Experimental geoneutrino spectra are usually presented in terms of number of events $N_i$ observed in bins of prompt
energy (or a related variable). The total number of events $N$ corresponds to an event rate $R$ at the detector via
\begin{equation}
N=\varepsilon\, T\,R\ ,
\end{equation}
where $\varepsilon$ is the detection efficiency, and $T$ is target exposure in units of 10$^{32}$ protons$\times$year,
with $R$ expressed in TNU.

\begin{figure}[t]
\centering
\epsfig{figure=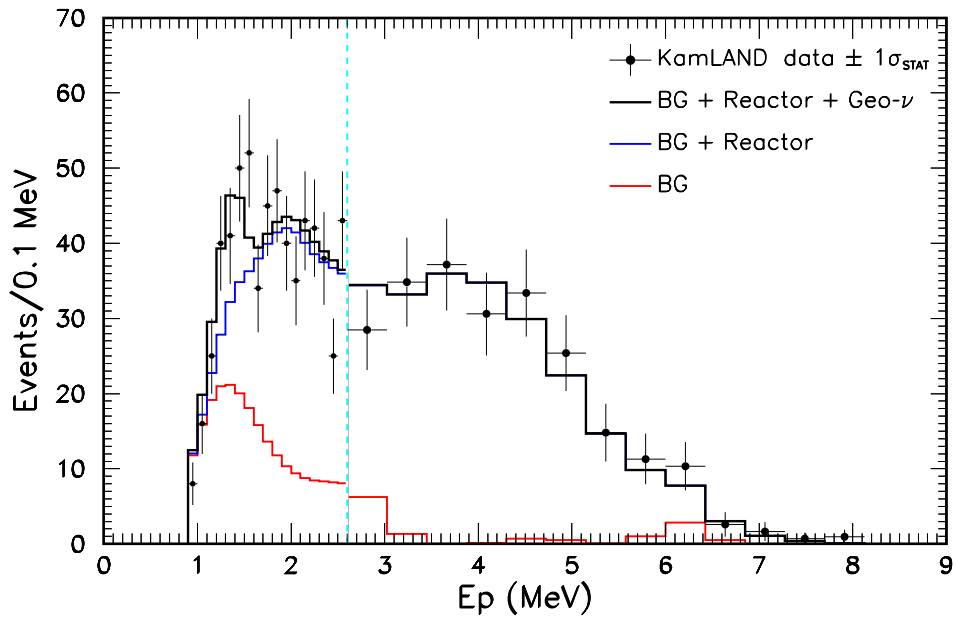,width=0.7\columnwidth}
\caption{KamLAND event spectrum as a function of the observed prompt energy $E_p$. Data points and backround (BG)
estimates are taken
from \protect\cite{KL08,Enom}; note the 
narrower binning in the geoneutrino energy range $E_p< 2.6$~MeV \protect\cite{Enom}.
The histogram represents our best-fit spectrum, with cumulative contributions
from background, plus reactor, plus geoneutrino events. 
\label{fig_02}}
\end{figure}

We use the latest KamLAND spectral data after an exposure 
$T=2.44\times 10^{32}$~protons$\times$year \cite{KL08}, with small-width bins in the geoneutrino range $E_p< 2.6$~MeV 
\cite{Enom}. We include the energy-dependent function $\varepsilon(E_p)$ from \cite{KL08}, which 
implies average detection efficiencies  
$\varepsilon_\mathrm{Th}=0.69$ and $\varepsilon_\mathrm{U}=0.78$ for Th and U events, respectively \cite{Enom}.
A Poisson $\chi^2$ function is constructed
\cite{Prev1} to account for statistical fluctuations. We introduce five systematic pulls,
one for the energy scale uncertainty, and four for   
the normalization of: $(i)$ reactor events;  $(ii)$ all events; $(iii)$  
ground-state and $(iv)$ excited-state contributions to the $^{13}$C$(\alpha,n)^{16}$O background.

Figure~2 shows the experimental KamLAND spectrum in terms of prompt energy, 
as well as the partial and total contributions to the
theoretical spectrum (background, reactor signal, and geoneutrino signal) at 
the best fit in the parameter space (\ref{PAR}). The geoneutrino signal is more pronounced
at low energies, consistently with a relatively large contribution from Th decay ($E_p\lesssim 1.7$~MeV).
Our geoneutrino fit results are very similar to the official ones \cite{KL08}; moreover, we reproduce the 
fit of \cite{Enom} in terms of total $\mathrm{Th}+\mathrm{U}$ rate versus $\mathrm{Th}-\mathrm{U}$ rate asymmetry (not shown).
We find that the null hypothesis (no geoneutrino signal in KamLAND) is rejected at $2.9\sigma$.

\begin{figure}[b]
\centering
\epsfig{figure=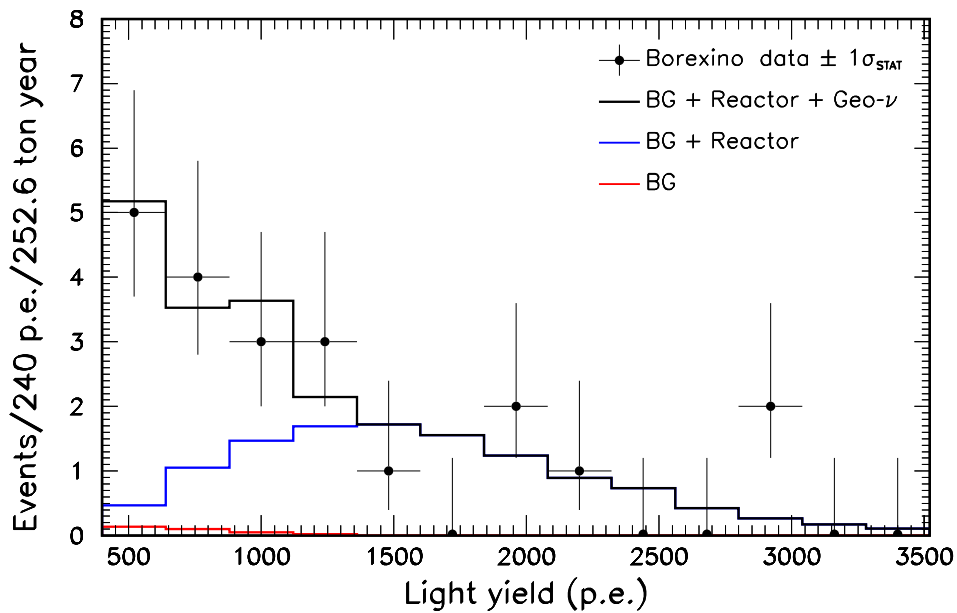,width=0.7\columnwidth}
\caption{Borexino event spectrum as a function of the light yield for positron events (p.e.). Data points
and background (BG) estimates are taken from \protect\cite{BX10}. 
The histogram represents our best-fit spectrum, with cumulative contributions
from background, plus reactor, plus geoneutrino events. 
\label{fig_03}}
\end{figure}

Concerning Borexino, we use the spectral data for an effective
exposure $\varepsilon\, T=0.152\times 10^{32}$~protons$\times$year, and in terms
of the light yield $Y$ for positron events---approximately 
equal to $Y\simeq 500\times E_p$/MeV \cite{BX10}. The data from the americium-beryllium source 
calibration in \cite{BX10} allow us to improve this approximation for $Y$, as well as to infer the energy resolution
width in Borexino.  
Our Poisson $\chi^2$ statistics includes two systematic pulls
for the normalization of all events and of reactor-only events. Since the background
is very small in Borexino, its systematic uncertainties are negligibile for our purposes.

Figure~3 shows the experimental Borexino spectrum in terms of light yield, as 
well as the separate contributions to our best-fit spectrum, in analogy with Fig.~2. Note the clear geoneutrino signal, which
covers the whole expected range $E_p\in [0.9,\,2.6]$~MeV ($Y\in[450,\,1300] $), consistent with
a leading contribution 
from U decay. Also in this case, we are able to reproduce quite well the
official geoneutrino fit results and plots of \cite{BX10} (not shown). 
We find that the null hypothesis in Borexino is rejected at $4.1\sigma$.

\subsection{Analyses with 4, 3, 2, and 1 degrees of freedom}

\begin{table*}[t]
\caption{\label{DOF}  Summary of adopted degrees of freedom and constraints.}
\begin{ruledtabular}
\begin{tabular}{clcccc}
$N_D$ & Constraints   	&  $R(\mathrm{Th+U})_\mathrm{KL}$ 
							&  $(\mathrm{Th}/\mathrm{U})_\mathrm{KL}$ 
							&  $R(\mathrm{Th+U})_\mathrm{BX}$
							&  $(\mathrm{Th}/\mathrm{U})_\mathrm{BX}$\\
\hline
4 & None & free & free & free & free \\
3 & $(\mathrm{Th}/\mathrm{U})_\mathrm{BX}=(\mathrm{Th}/\mathrm{U})_\mathrm{KL}$ & free & free & free & --- \\
2 & $(\mathrm{Th}/\mathrm{U})_\mathrm{BX}=(\mathrm{Th}/\mathrm{U})_\mathrm{KL}$
    and $R_\mathrm{BX}=1.15\,R_\mathrm{KL}$ & free & free & --- & --- \\
1 & $(\mathrm{Th}/\mathrm{U})_\mathrm{BX}=(\mathrm{Th}/\mathrm{U})_\mathrm{KL}=3.9$ and $R_\mathrm{BX}=1.15\,R_\mathrm{KL}$ 
  & free & --- & --- & ---  
\end{tabular}
\end{ruledtabular}
\end{table*}

After marginalization of the oscillation parameters in (\ref{PAR}), the most general geoneutrino parameter space is spanned 
by four event rates. We find it useful to define four equivalent geoneutrino degrees of freedom 
($N_D=4$), namely, the total rate and the Th/U ratio
probed by KamLAND and Borexino, 
\begin{eqnarray}
R(\mathrm{Th}+\mathrm{U})_\mathrm{KL}&=&R(\mathrm{Th})_\mathrm{KL}+R(\mathrm{U})_\mathrm{KL}\ ,\\
(\mathrm{Th}/\mathrm{U})_\mathrm{KL}&=& \left [R(\mathrm{Th})_\mathrm{KL}/R(\mathrm{U})_\mathrm{KL}\right]/6.96\times 10^{-2}\ ,\\
R(\mathrm{Th}+\mathrm{U})_\mathrm{BX}&=&R(\mathrm{Th})_\mathrm{BX}+R(\mathrm{U})_\mathrm{BX}\ ,\\
(\mathrm{Th}/\mathrm{U})_\mathrm{BX}&=& \left [R(\mathrm{Th})_\mathrm{BX}/R(\mathrm{U})_\mathrm{BX}\right]/6.96\times 10^{-2}\ .
\end{eqnarray}
One can reduce the degrees of freedom to $N_D=3$ by assuming
that KL and BX probe essentially the same Th/U ratio,
\begin{equation}
\label{SAME}
(\mathrm{Th}/\mathrm{U})_\mathrm{BX} = (\mathrm{Th}/\mathrm{U})_\mathrm{KL}\ .
\end{equation}
If, in addition, the scaling law in Eq.~(\ref{SCALE}) is assumed, then $N_D=2$. Finally, if the chondritic
Th/U estimate in~(\ref{CHO}) is also assumed, then $N_D=1$. These four options, involving
an increasing Earth model dependence
for decreasing $N_D$, are summarized in Table~\ref{DOF}.

A final remark is in order. As discussed in the next Section, the allowed ranges for the KamLAND and Borexino geoneutrino 
degrees of freedom may extend beyond plausible expectations, 
where the constraints in Eqs.~(\ref{CHO}),~(\ref{SCALE}) and (\ref{SAME}) are not really 
justified by any Earth model.
Therefore, while the analysis for $N_D=4$ is completely general, the results of
constrained analyses ($N_D\leq 3$) must be taken with a grain of salt.

\subsection{Analysis with an additional degree of freedom: the georeactor}

It has been proposed \cite{He96} that there could be enough uranium in the Earth's core to
naturally start a nuclear fission chain over geological timescales, with a typical power (at
the current epoch) of $P_\mathrm{geo}\simeq 3$-Ð10 TW \cite{He05}. This hypothesis is  disfavored by 
various geochemical and geophysical arguments \cite{NoRe}. Particle physics offers an independent 
probe of the hypothesis, since a georeactor 
would alter the observable energy (and time) spectra of $\overline\nu_e$ events \cite{Ragh}. In particular, we reported
in \cite{Prev2}  an analysis of earlier KamLAND data in the energy and time domain, providing
an upper bound $P_\mathrm{geo}\lesssim 13$~TW at $95\%$~C.L. Further KamLAND and Borexino data have reduced
the upper bound to $\lesssim 6.2$~TW at 90\% C.L.\ \cite{KL08} and $\lesssim 3$~TW at 95\% C.L.\ \cite{BX10}, respectively.

Here we update our previous analysis \cite{Prev2}, by assuming a contribution from a georeactor at the center of the Earth
(with unconstrained $P_\mathrm{geo}$) in the KamLAND and Borexino energy spectra, for each 
of the four cases in Table~\ref{DOF}. With respect to \cite{Prev2}, the current analysis does not include
the event time information, which has not been released by the experiments \cite{KL08,BX10}.

\section{Results}

In this Section we describe the results of our analysis, in terms of both joint and separate bounds on 
the Th/U and $R(\mathrm{Th}+\mathrm{U})$ variables.

\subsection{Joint $1\sigma$ regions for Th/U and {\boldmath $R(\mathrm{Th}+\mathrm{U})$}}

Figure~4 shows the  $1\sigma$ contours ($\Delta\chi^2=1$) in the plane charted
by the total event rate $R(\mathrm{Th}+\mathrm{U})$ and by the mass abundance ratio Th/U for KamLAND
and Borexino. The degrees of freedom decrease from $N_D=4$ to $N_D=1$ from top to bottom,
according to the constraints in Table~\ref{DOF}.

\begin{figure}[t]
\centering
\epsfig{figure=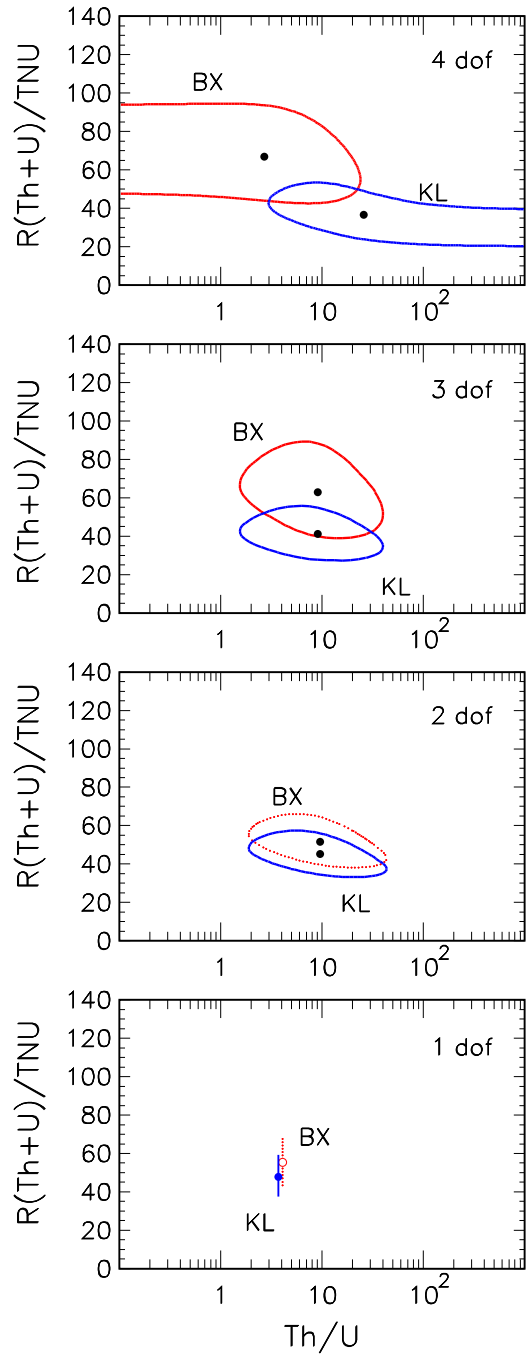,width=0.38\columnwidth}
\caption{KamLAND (KL) and Borexino (BX) geoneutrino analysis 
in the plane charted by the total rate $R(\mathrm{Th}+\mathrm{U})$ and by the
mass abundance ratio Th/U. The curves represents  $1\sigma$ contours ($\Delta\chi^2=1$) around 
the best-fit points (thick dots).
From top to bottom,
the degrees of freedom decrease from $N_D=4$ to $N_D=1$, as reported in Table~\protect\ref{DOF}.
\label{fig_04}}
\end{figure}

The upper panel ($N_D=4$) shows that both KamLAND and Borexino place upper and lower
bounds to the total event rate $R(\mathrm{Th}+\mathrm{U})$. These bounds are consistent with typical 
Earth model expectations, which
span the $1\sigma$ range 29--41 TNU for KamLAND and 34--48 TNU for Borexino (see Table~11 in \cite{Fi07}). 
However, neither KamLAND nor Borexino can currently determine Th and U separately. In particular,
KamLAND is compatible with all events being from Th decay (Th/U$=\infty$), while
Borexino is compatible with all events being from U decay (Th/U$=0$), as anticipated in the
context of Figs.~2 and 3, respectively. However, a broad range of Th/U values 
appears to be compatible with both
KamLAND and Borexino results at $1\sigma$; this range excludes the extreme
cases with null Th or U signal, and includes the chondritic value Th/U$=3.9$. 
Therefore, it makes sense to reduce the degrees of freedom by imposing that
the two experiments probe the same Th/U ratio as in Eq.~(\ref{SAME}).

The second panel ($N_D=3$) shows the results of such exercise, providing both upper 
and lower $1\sigma$ limits on the Th/U ratio, with a best
fit which is only a factor of $\sim\! 2$ higher than the chondritic value.  
The total rate estimates are not significantly altered with respect to the case with $N_D=4$.
Therefore, under the rather general assumption in Eq.~(\ref{SAME}), the combination
of KamLAND and Borexino data starts to be sensitive to the global Th/U ratio of the Earth,
although only at the $\sim\! 1\sigma$ level; as discussed below, current Th/U constraints
disappear at $\sim\! 1.5\sigma$.

The results in the third panel ($N_D=2$) include, in addition, the approximate scaling
assumption in Eq.~(\ref{SCALE}). In this case, the KamLAND parameters $R(\mathrm{Th}+\mathrm{U})$
and Th/U are conventionally taken as free, while the corresponding Borexino parameters are derived
(hence the ``dotted'' BX contour in the panel).
In this case, the best fits for the total rates
are located slightly above the quoted Earth model expectations (29--41 TNU for KamLAND 
and 34--48 TNU for Borexino \cite{Fi07}), with $1\sigma$ uncertainties at the level of $\sim 30\%$, 
dominated by KamLAND data. Concerning Th/U, the best fits and $1\sigma$ ranges 
are not significantly altered with respect to the previous case with $N_D=3$.

The comparison of the three panels with $N_D=4$, 3 and 2 shows that the current
constraints on the total rates and on the Th/U ratio are approximately independent: they do not 
significantly affect each other within present data. The weak negative correlation 
in the $1\sigma$ contours reflects the fact the overall rate $R(\mathrm{Th}+\mathrm{U})$ 
increases somewhat faster for larger U contribution as compared to Th contribution, 
the latter being confined at low energy (see the spectra in Fig.~1).

Finally, the results in the lower panel of Fig.~4 ($N_D=1$) include, in addition to the previous
constraints, the chondritic  estimate in~(\ref{CHO}).%
\footnote{In the panel, the Th/U coordinates of KL and BX are slightly displaced from 3.9 for the sake of clarity.}
In this case, the estimated KL total rate is $R(\mathrm{Th}+\mathrm{U})_\mathrm{KL}=47.7\pm 11.2$~TNU, 
the BX total rate being a factor $\sim\!1.15$ higher by construction,  $R(\mathrm{Th}+\mathrm{U})_\mathrm{BX}=54.9\pm 12.9$~TNU. 
These results show a preference for Earth models with relatively high expectations in Th and U contents, although
within very large uncertainties at present.
We report in Table~\ref{SIGMA} a numerical summary of the $1\sigma$ ranges for 
the total rate and Th/U ratio, in each of the four cases considered.

In the most constrained case ($N_D=1$), where the total rate error   
is reduced to $\sim\! 23\%$, it makes sense to infer indications about 
the associated radiogenic heat $H(\mathrm{Th}+\mathrm{U})$ via the approximate 
$(H,\,R)$ correlation in Eq.~(\ref{HR}). By fixing the KL rate at its central value,
$R(\mathrm{Th}+\mathrm{U})_\mathrm{KL}=47.7$~TNU, one would obtain an allowed range 
$H(\mathrm{Th}+\mathrm{U})\simeq 21$--35~TW, somewhat above the plausible expectations
of 14--18~TW (see Fig.~23 in \cite{Fi07}). However, including the $1\sigma$
rate uncertainties, the allowed range in significantly enlarged: $H(\mathrm{Th}+\mathrm{U})\simeq 10$--49~TW.
The upper value is not particularly meaningful, being larger than the ``fully radiogenic'' limit of $\sim\! 40$~TW. 
The lower value of $\sim\! 10$~TW, however, exceeds the ``guaranteed'' contribution from Th and U in the crust 
($\sim\! 6$~TW \cite{Fi07}),
and suggests, indirectly, the presence of an additional contribution from a different 
reservoir---which can be naturally identified with the mantle. 

In conclusion, the combination of KamLAND and Borexino data brings to surface some 
intriguing---although still statistically weak---pieces of information: 
($i$) preferred Th/U values in broad agreement
with chondritic expectations; ($ii$) slight preference for relatively high Th and U contents in the
Earth; and ($iii$) hints of a mantle contribution to the total geoneutrino signal. 
We remark that these indications emerge only at the $\sim \!1\sigma$ level from the current,
low-statistics data samples.

\begin{table*}[t]
\caption{\label{SIGMA} Best fits and $1\sigma$ ranges from the data analysis with degrees of freedom $N_D\leq 4$.
Event rates $R$ are expressed in TNU.
Derived or fixed numbers are given in brackets.}
\begin{ruledtabular}
\begin{tabular}{ccccc}
$N_D$ 	                 	&  $R(\mathrm{Th+U})_\mathrm{KL}$
							&  $(\mathrm{Th}/\mathrm{U})_\mathrm{KL}$ 
							&  $R(\mathrm{Th+U})_\mathrm{BX}$
							&  $(\mathrm{Th}/\mathrm{U})_\mathrm{BX}$\\
\hline
4		&$36.8^{+16.2}_{-16.1}$		&$25.9^{+\infty}_{-22.9}$		&$66.9^{+27.3}_{-23.8}$		&$2.7^{+20.2}_{-2.7}$  \\
3		&$41.3^{+14.0}_{-12.6}$		&$9.1^{+23.5}_{-7.4}$          &$63.0^{+26.0}_{-24.0}$    &$\left[9.1^{+23.5}_{-7.4}\right]$  \\
2		&$45.1^{+11.8}_{-11.2}$		&$9.6^{+33.7}_{-7.6}$          &$\left[51.7^{+13.6}_{-12.9}\right]$   &$\left[9.6^{+33.7}_{-7.6}\right]$  \\
1		&$47.7^{+11.2}_{-11.2}$		&$[3.9]$                       &$\left[54.9^{+12.9}_{-12.9}\right]$   &$[3.9]$ \\
\end{tabular}
\end{ruledtabular}
\end{table*}

\subsection{Separate bounds on Th/U and {\boldmath $R(\mathrm{Th}+\mathrm{U})$}}

In this section we discuss the separate projections of the previous results onto the
variables Th/U and $R(\mathrm{Th}+\mathrm{U})$, in terms of 
standard deviations from their best fit ($N_\sigma=\sqrt{\Delta\chi^2}$).

\begin{figure}[t]
\centering
\epsfig{figure=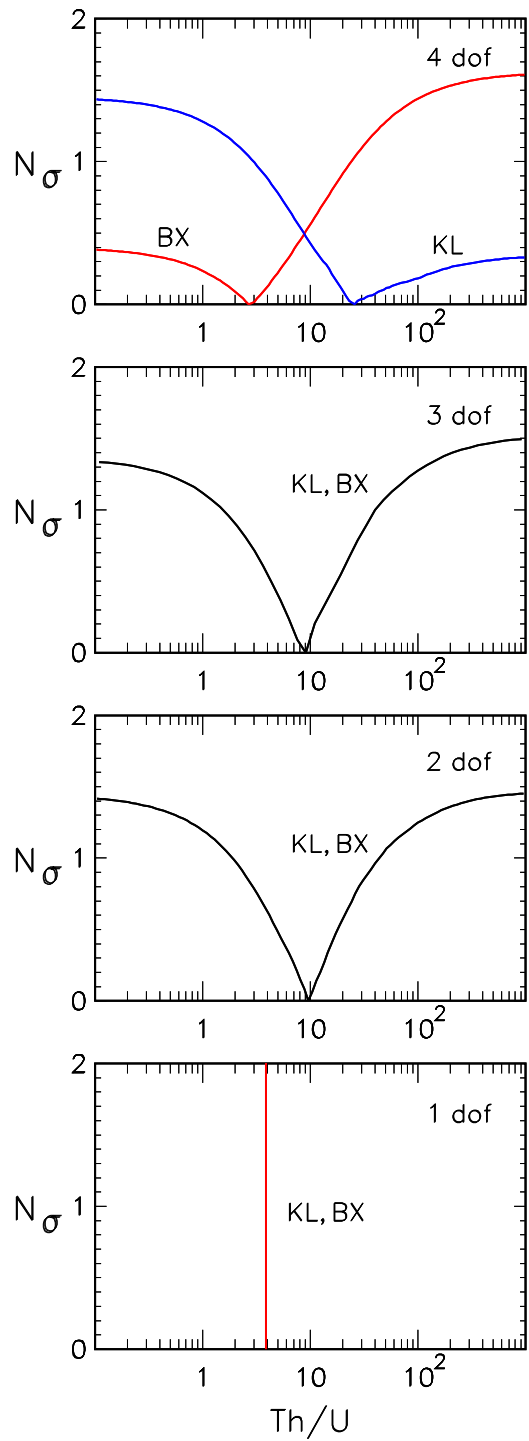,width=0.38\columnwidth}
\caption{
KamLAND and Borexino analysis: constraints on the Th/U ratio
in terms of standard deviations, $N_\sigma=\sqrt{\Delta \chi^2}$.
\label{fig_05}}
\end{figure}

Figure~5 shows the constraints on the Th/U ratio from KamLAND and Borexino. In 
the upper panel $(N_D=4)$, one can see at a glance that KamLAND and Borexino
data place lower and upper $1\sigma$ limits on Th/U, respectively, but have
no significant sensitivity in the opposite directions. The bounds are still
statistically weak, as they vanish at $\sim\! 1.3\sigma$ in KamLAND
and at $\sim\! 1.6\sigma$ in Borexino. The middle panels show that 
KamLAND and Borexino provide joint limits on Th/U at the $1\sigma$ level,
with no significant variation between the cases with $N_D=3$ and $N_D=2$.
The lower panel is just a ``Dirac delta'' at Th/U$=3.9$.

\begin{figure}[t]
\centering
\epsfig{figure=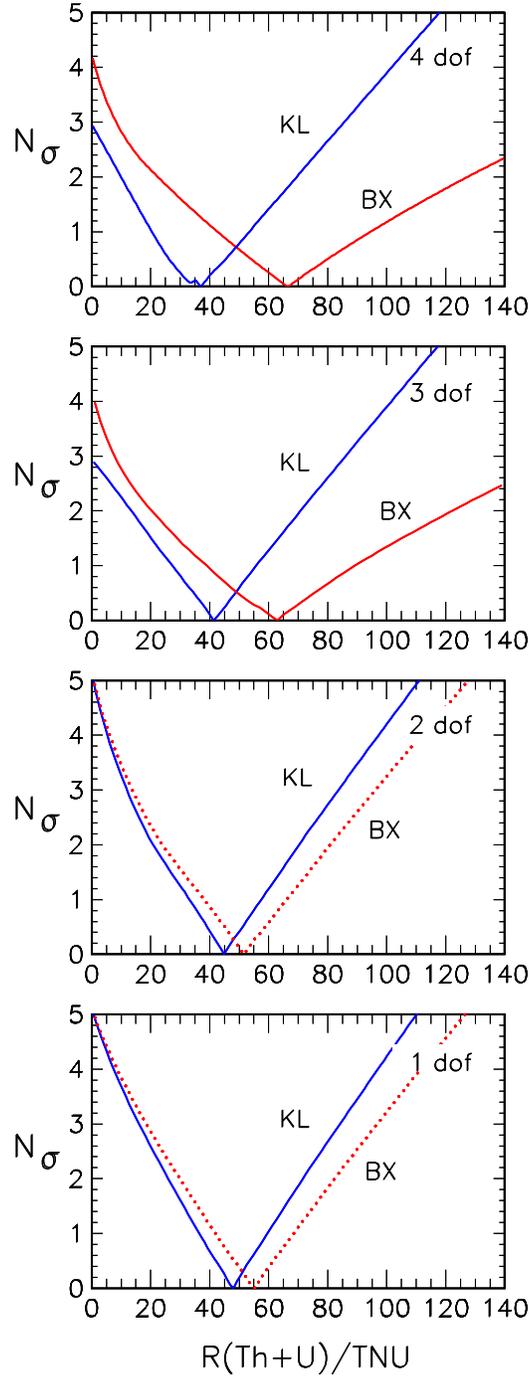,width=0.38\columnwidth}
\caption{KamLAND and Borexino analysis: constraints on the total rate $R(\mathrm{Th}+\mathrm{U})$
in terms of standard deviations $N_\sigma$.
\label{fig_06}}
\end{figure}

Figure~6 shows the constraints on the total event rate $R(\mathrm{Th}+\mathrm{U})$
in KamLAND and Borexino. The two upper panels shows that the null hypothesis 
of no geoneutrino signal is  rejected at $2.9\sigma$ in KamLAND and at 
$4.1\sigma$ in Borexino, almost independently of
the Th/U constraint in Eq.~(\ref{SAME}). The two lower panels show
the significant error reduction induced by the 
scaling law  assumption in Eq.~(\ref{SCALE}), which strengthens the
null hypothesis rejection at the $5\sigma$ level. By comparing the two lower
panels, it appears that the chondritic assumption in~(\ref{CHO}) has
a minor impact on combined rate constraints. 

In conclusion, current indications on the Th/U ratio emerge only at the level
of $\sim 1\sigma$, while global geoneutrino signals emerge at $\gtrsim 3\sigma$,
reaching $\sim\!5\sigma$ in constrained combinations. Bounds on Th+U rates and on
the Th/U ratio
are largely uncorrelated. Detailed $1\sigma$ ranges
for both free and derived parameters are reported in Table~\ref{SIGMA}.

\subsection{Bounds on the power of a hypothetical georeactor}

We have repeated the various analyses by including 
a spectral contribution from a hypothetical georeactor with unknown power $P_\mathrm{geo}$, located
at the center of the Earth.
The results disfavor this hypothesis in all cases, if $P_\mathrm{geo}$ exceeds a few TW.
In the most general case (unconstrained geoneutrino and georeactor event rates), we 
find that KamLAND and Borexino place the $2\sigma$ upper limits $P_\mathrm{geo}\lesssim 4.1$~TW
and $\lesssim 6.7$~TW, respectively.   
In combination,  the joint KamLAND+Borexino limit reads 
\begin{equation}
P_\mathrm{geo}\lesssim 3.9~\mathrm{TW\ at\ } 2\sigma  \ 
(\lesssim 5.2~\mathrm{TW\ at\ } 3\sigma)\ , 
\end{equation}
almost independently of the chosen geoneutrino degrees of freedom. This is to be expected,
since the georeactor spectrum extends well above the geoneutrino energy range.

Our combined constraints on $P_\mathrm{geo}$ appear to be dominated by Borexino data, but do not improve upon the
official Borexino limit (quoted as $P_\mathrm{geo}\lesssim 3$~TW at $2\sigma$ \cite{BX10}), presumably because 
we cannot include information in the time domain, which is currently unpublished. Our analysis of older
KamLAND data \cite{Prev2} has indeed shown that the time spectra can add significant
constraints to $P_\mathrm{geo}$.

\section{Conclusions}

We have performed a detailed analysis of current geoneutrino events 
from Th and U decay chains as detected in KamLAND \cite{KL08} and Borexino \cite{BX10},
within the more general context of low-energy neutrino oscillation data from long-baseline
reactor and solar sources, and of a broad range of Earth model expectations taken from \cite{Fi07}.

The relevant parameter space is spanned by the total Th+U event rates  
and the Th/U ratio in the two experiments, while 
the oscillation parameters $(\delta m^2,\,\theta_{12},\,\theta_{13})$
are  marginalized away. Constrained analyses with fewer
degrees of freedom are obtained by successively assuming for both experiments
a common Th/U ratio [Eq.~(\ref{SAME})], 
a common scaling of $\mathrm{Th}+\mathrm{U}$ event rates [Eq.~(\ref{SCALE})], and a fixed (chondritic) Th/U value
[Eq.~(\ref{CHO})]. Cases with a hypothetical georeactor, involving an additional degree of freedom,
are also considered.

The results are in agreement with typical Earth model expectations, although within still
large uncertainties.
The global Th+U geoneutrino signal emerges at $2.9\sigma$ and $4.1\sigma$ in KamLAND
and Borexino, respectively, and can reach the overall $5\sigma$ level in 
combination. The data disfavor the hypothesis of a georeactor, and limit its power
to $P_\mathrm{geo}\lesssim 3.9$~TW at  $2\sigma$  (or $P_\mathrm{geo}\lesssim 5.2$~TW at $3\sigma)$.

Weaker---but potentially interesting---pieces of information emerge
at the $\sim 1\sigma$ level, including: 
($i$) preferred Th/U values in broad agreement
with chondritic expectations; ($ii$) slight preference for relatively high Th and U contents in the
Earth; and ($iii$) possible hints of a (mantle) contribution in excess
of the ``guaranteed'' signal from the crust.  
Significantly higher statistics, possibly from new large-volume detectors and with
some directional sensitivity, will be needed
to promote these intriguing indications to more robust signals improving our
understanding of the Earth's interior.

\acknowledgments

E.L.\ thanks G.~Fiorentini, A.~Ianni and W.F.~McDonough for many useful discussions about geoneutrino physics in the
last few years.  The work of G.L.F., E.L.\ and A.M.R.\ is partly supported by the Italian
Ministero dell'Istruzione, dell'Universit\`a e della Ricerca (MIUR) through the
Progetto di Rilevante Interesse Nazionale (PRIN) ``Fisica Astroparticellare: Neutrino ed
Universo Primordiale,'' and partly by the 
 Istituto Nazionale di Fisica Nucleare (INFN) through the  research initiative   
 ``Fisica Astroparticellare FA51.'' 
 The work of A.P.\ is supported by the Spanish Ministerio de Educaci\'on y Ciencia (MEC) under the I3P program,
by the Spanish Grants FPA2008-00319, CSD2009-00064, PROMETEO/2009/091,
and by the European network UNILHC (PITN-GA-2009-237920).


\newpage
\begin{thebibliography}{99}

\bibitem{Fi07}
  G.~Fiorentini, M.~Lissia and F.~Mantovani,
  ``Geo-neutrinos and Earth's interior,''
  Phys.\ Rept.\  {\bf 453}, 117 (2007).
  
\bibitem{Mc08}  
 W.F.~McDonough and R.~Arevalo Jr.,
  ``Uncertainties in the composition of Earth, its core and silicate sphere,''
 Proceedings on {\em Neutrino 2008}, XXIII International Conference on Neutrino Physics and Astrophysics
 (Christchurch, New Zealand, 2008), ed.\ by J.~Adams, F.~Halzen, and S.~Parke, 
 Journal of Physics Conference Series {\bf 136}, 022006 (2008).
 
\bibitem{Fo06}
  G.~L.~Fogli, E.~Lisi, A.~Marrone and A.~Palazzo,
  ``Global analysis of three-flavor neutrino masses and mixings,''
  Prog.\ Part.\ Nucl.\ Phys.\  {\bf 57}, 742 (2006).
  
\bibitem{KL05} KamLAND Collaboration, T.~Araki {\em et al.}, 
``Experimental investigation of geologically produced antineutrinos with KamLAND,''  
Nature {\bf 436}, 499 (2005).

\bibitem{KL08} KamLAND Collaboration, S.~Abe {\em et al.},
``Precision Measurement of Neutrino Oscillation Parameters with KamLAND,''
Phys.\ Rev.\ Lett.\ {\bf 100}, 221803 (2008).
  
\bibitem{BX10} Borexino Collaboration, G.~Bellini {\em et al.},
``Observation of Geo-Neutrinos,''
Phys.\ Lett.\ B {\bf 687}, 299 (2010).

\bibitem{Mc09} R.~Arevalo Jr, W.F.~McDonough, and M.~Luong, 
``The K/U ratio of the silicate Earth: Insights into mantle composition, structure and
thermal evolution,'' Earth Planet.\ Sci.\ Lett.\ {\bf 278}, 361 (2009).

\bibitem{Po93}	H.N.~Pollack, S.J.~Hurter, and J.R.~Johnson, 
				``Heat Flow from the Earth's Interior: Analysis of the Global Data Set,''
				Rev.\ Geophys.\ {\bf 31}, 267--280 (1993).

\bibitem{Ja07}		C.~Jaupart, S.~Labrosse, and J.-C. Mareschal, 
 				``Temperatures, Heat and Energy in the Mantle of the Earth,''
				 in {\em Treatise on Geophysics}, Vol.~7, pp.~253--303, edited by G.~Schubert
               (Elsevier, Oxford, 2007).
 
\bibitem{Fo07} G.L.\ Fogli,  E. Lisi, A.\ Palazzo, and A.M.\ Rotunno,
``Geoneutrinos: A systematic approach to uncertainties and correlations,''
Proceedings of {\em Neutrino Geophysics\/} (Honolulu, Hawaii, 2005), edited by S.T.\ Dye,
Earth, Moon, and Planets {\bf 99}, 111 (2007).


\bibitem{He96} J.M.~Herndon, ``Sub-structure of the inner core of the Earth,''
Proc.\ Natl.\ Acad.\ Sci.\ U.S.A.\ {\bf 93}(2), 646 (1996); 
``Nuclear georeactor origin of oceanic basalt $^3$He/$^4$He, evidence, and implications,'' 
ibidem {\bf 100}(6), 3047 (2003).



\bibitem{Isot} Isotopes Project database, available at {\tt ie.lbl.gov}.


\bibitem{BeJa} H.\ Behrens and J.\ Janecke, 
   ``Numerical tables for beta decay and electron capture,'' 
   (Springer Verlag, Berlin, 1969).
   
\bibitem{StVi}
  A.~Strumia and F.~Vissani,
  ``Precise quasielastic neutrino nucleon cross section,''
  Phys.\ Lett.\  B {\bf 564}, 42 (2003).

\bibitem{Prev1}
  G.L.~Fogli, E.~Lisi, A.~Marrone, D.~Montanino, A.~Palazzo and A.M.~Rotunno,
  ``Solar neutrino oscillation parameters after first KamLAND results,''
  Phys.\ Rev.\  D {\bf 67}, 073002 (2003).
  
\bibitem{Prev2}  
  G.L.~Fogli, E.~Lisi, A.~Palazzo and A.M.~Rotunno,
  ``KamLAND neutrino spectra in energy and time: Indications for reactor  power
  variations and constraints on the georeactor,''
  Phys.\ Lett.\  B {\bf 623}, 80 (2005).

\bibitem{Prev3}
 G.L.\ Fogli, E.\ Lisi, A.\ Marrone, A.\ Melchiorri, A.\ Palazzo, A.M.\ Rotunno, P.\ Serra, J.\ Silk, and A.\ Slosar, 
 ``Observables sensitive to absolute neutrino masses. II,''
 Phys.\ Rev.\ D {\bf 78}, 033010 (2008).

\bibitem{Nucl} Databases of nuclear power plants are available at: 
International Nuclear Safety Center 					({\tt www.insc.anl.gov}); 
Global Resource Information Database (GRID)---Europe 	({\tt www.grid.unep.ch}); 
International Atomic Energy Agency 						({\tt www.iaea.org}); 
World Nuclear Association 					            ({\tt www.world-nuclear.org}).

\bibitem{Hint}
  G.L.~Fogli, E.~Lisi, A.~Marrone, A.~Palazzo and A.M.~Rotunno,
  ``Hints of $\theta_{13}>0$ from global neutrino data analysis,''
  Phys.\ Rev.\ Lett.\  {\bf 101}, 141801 (2008).


\bibitem{SAGE}
  SAGE Collaboration, J.N.~Abdurashitov {\it et al.},  
  ``Measurement of the solar neutrino capture rate with gallium metal. III:
  Results for the 2002--2007 data-taking period,''
  Phys.\ Rev.\  C {\bf 80}, 015807 (2009).


\bibitem{GALL}  
F.~Kaether, W.~Hampel, G.~Heusser, J.~Kiko and T.~Kirsten,
  ``Reanalysis of the GALLEX solar neutrino flux and source experiments,''
  Phys.\ Lett.\  B {\bf 685}, 47 (2010).

\bibitem{LETA}
  SNO Collaboration, B.~Aharmim {\it et al.},
  ``Low Energy Threshold Analysis of the Phase I and Phase II Data Sets of the
  Sudbury Neutrino Observatory,''
  arXiv:0910.2984 [nucl-ex].

\bibitem{Sere}
  A.~Serenelli, S.~Basu, J.W.~Ferguson and M.~Asplund,
  ``New Solar Composition: The Problem With Solar Models Revisited,''
  arXiv:0909.2668 [astro-ph.SR].


\bibitem{Pull} 
G.L.~Fogli, E.~Lisi, A.~Marrone, D.~Montanino and A.~Palazzo,
  ``Getting the most from the statistical analysis of solar neutrino
  oscillations,''
  Phys.\ Rev.\  D {\bf 66}, 053010 (2002).


\bibitem{Enom} S.\ Enomoto, ``KamLAND Geoneutrino Studies,'' talk at the Workshop {\em Neutrino Geoscience 2008\/}
 (Sudbury, Ontario, Canada, 2008); available at {\tt geonu.snolab.ca}.

\bibitem{He05} J.M.~Herndon and D.A.~Edgerley,  arXiv:hep-ph/0501216.

\bibitem{Ragh}
  R.S.~Raghavan,
  ``Detecting a nuclear fission reactor at the center of the earth,''
  arXiv:hep-ex/0208038.

\bibitem{NoRe} W.F.~McDonough, ``Compositional Models for the Earth's Core,'' in {\em Treatise on Geochemistry},
Vol. II, edited by R.W.~Carlson (Elsevier-Pergamon, Oxford, 2003).



\end{thebibliography}
\end{document}